\documentclass[aps,pre,floats,twocolumn,floatfix,footinbib]{revtex4-1}
\usepackage{amsmath,amsfonts,amssymb,stmaryrd,amsbsy,bm}
\usepackage{mathtools}
\usepackage{graphicx}
\usepackage{breakurl}
\usepackage{hyperref}

\makeatletter
\def\squiggly{\bgroup \markoverwith{\textcolor{red}{\lower3.5\p@\hbox{\sixly \char58}}}\ULon}
\makeatother

\newcommand{\dom}{\mathrm{dom}\,}
\newcommand{\defeq}{ {\kern 0.2em}:{\kern -0.5em}={\kern 0.2 em} }  
\newcommand{\eqdef}{ {\kern 0.2em}={\kern -0.5em}:{\kern 0.2 em} }  
\newcommand{\Expct}{\Bbb E}
\newcommand{\expct}[2][]{\left\langle {#2} \right\rangle_{#1}}

\newcommand{\Gen}{{\cal Z}}
\newcommand{\Genint}{{\cal Z}_{ \mathrm{fst} }}

\newcommand{\con}[1]{\overline{#1}}
\DeclareMathOperator{\Sym}{\mathrm{Sym}}
\newcommand{\Pfst}{P_{\text{fst}}}
\newcommand{\efst}[1]{\expct[\text{fst}]{#1}}
\newcommand{\Arr}[3]{{#1}\colon {#2} \rightarrow {#3}} 
\newcommand{\Reals}{{\mathbb R}}    
\newcommand{\bigO}[1]{{\cal O}\left({#1}\right)}
\newcommand{\transpose}[1]{{{#1}^\top}}
\newcommand{\evec}[1]{{\bm e}_{#1}}  

\DeclareMathOperator{\DIV}{\mathrm{Div}}

\begin{document}
\title{Drift-diffusion processes from elimination of fast variables \\
  under inhomogeneous conditions}
\author{Paul E. Lammert}
\email{lammert@psu.edu}
\affiliation{Department of Physics, Pennsylvania State University, University Park, PA 16802}

\begin{abstract}
  The problem of eliminating fast-relaxing variables to obtain an effective
  drift-diffusion process in position is solved in a uniform and
  straightforward way
  for models with velocity a function jointly of position and fast variables.
  A more unified view is thereby obtained of the effect of environmental
  inhomgeneity on the motion of a diffusing particle, in particular,
  whether a drift is induced, covering both passive and active particles.
  Infinitesimal generators (equivalently, drift-diffusion fields) for the
  contracted
  processes are worked out in detail for several models.
\end{abstract}

\date{Nov. 1, 2019}
\maketitle

\section{Introduction}

Thirty years ago, van Kampen\cite{van-Kampen-88} investigated the effects of
environmental inhomogeneity in models of random motion of passive particles.
In general, the problem continues to attract
interest\cite{Christensen+Pedersen-03,Bringuier-09,Yang+Ripoll-13,Livshits-16,
  Bhattacharyay-19,Baldovin+19,Abdoli+19}, and has been especially revived
for active matter\cite{Hong+07,Butler+15,Byun+17,Zhao+18,Popescu+18}.
Cates and Tailleur\cite{Cates+Tailleur-13} and Vuijk {\it et alia}\cite{Vuijk+19},
in particular, studied variants of a model to be taken up here.
Consider, for instance, a micron-sized particle propelling itself via chemical reactions
using fuel from the environment, at a speed increasing with fuel concentration.
Random torques cause its direction of motion to wander (time scale $\sim 1$ sec),
so that its motion is diffusive on longer time scales.
Will a fuel concentration gradient cause drift?
If so, the mechanism would seem to be rather different than for a passive particle.

This paper presents a unified perspective for {\it velocity models} ---
models for which velocity is given as a function of continuous position and
fast-relaxing variables. I show how to eliminate the fast variables to obtain the
infinitesimal generator for a positional drift-diffusion in the
markovian limit by a uniform and natural method. Drift is just a part of
this {\it contraction} problem. Although the infinitesimal generator is interconvertible
with a Fokker-Planck-type equation, being a dual, or adjoint formulation,
the method given here
seems significantly simpler than methods which work with a probability density
from the beginning.
Significant generalizations of both the active particle model mentioned above
and the Langevin-Kramers model of a brownian particle discussed by van Kampen
are treated in detail, together with a coupling of them.
All those belong to the class of {\it quasilinear} velocity models, for which
practical formulas for drift-diffusion fields are given,
requiring a handful of tensor fields on position space expressing simple properties
of the original velocity model and its fast equilbria.
Intuition suggests that, all other things being equal, a passive brownian particle
drifts toward lower velocity damping while the simple active particle drifts toward
higher fuel concentration. 
Suitably embellished, these are the only two general mechanisms of gradient-induced
drift for quasilinear models, and the same is arguably roughly true for general
velocity models.

Zoom out now to situate that contraction problem in a broader idea of coarse-graining.
Drift-diffusions, characterized by drift ($u$) and diffusion ($D$) coefficients, and markovian
random processes with continuous sample paths, are essentially the same thing.
Therefore, to make a drift-diffusion approximation (of a simple random walk, for
an elementary example) is to approximate by a markovian process with continuous
sample paths. If the process as given lacks markovianity, it may display it approximately
over longer time scales --- a {\em temporal} coarse-graining is called for.
The processes treated in this paper have a velocity determined jointly by 
position and some {\it fast} variables, that is, with quickly decaying correlations.
They belong to this first general category.
If the process as given has discontinuous sample paths
(a so-called hopping
model\cite{van-Kampen-88,Van-Kampen-Book,Christensen+Pedersen-03,Smith+17},
for instance)
it might be approximable as continuous after a {\em spatial} coarse-graining.
Those are outside the purview of this paper.
A seeming third category comprises discrete time markovian processes, such as the
archetypal random walk.
However, viewed as {\em time-homogeneous} continuous-time processes,
as they must be to prepare for a drift-diffusion approximation, they are
nonmarkovian. Therefore, they too conform to the rule of temporal/spatial
coarse-graining to obtain markovianity/continuous paths.

Here is a content guide.
Section \ref{sec:Feller} reviews the aforementioned 
equivalence of {drift-diffusions} with {markovian with continuous paths},
and the observable-centered (``Heisenberg picture'') semigroup formalism
we will use.
Section \ref{sec:contraction} solves the {\it contraction} problem ---
elimination of fast variables --- for a general {\it velocity model},
leading to expressions for the infinitesimal generator of the contracted
process in terms of time correlation functions in fast-equilibrium.
[See Eqs. (\ref{eq:contracted-markovian-gen}), (\ref{eq:con-u}), (\ref{eq:con-D})].
Section \ref{sec:elaboration} elaborates the Langevin-Kramers model,
to one with arbitary tensor parameters, and a simple model of an active particle,
to one with fully anisotropic rotational diffusion characteristics,
as well as a coupling of the two.
The elaborated models are fit into the framework of ({\it quasilinear}) models
describable in terms of intrinsic drift ($\efst{v}$), decay rate ($\Lambda$),
velocity transformation ($\sigma$), and kinetic ($K$), tensor fields.
A general conclusion here is that drift which is not already implied by fast equilibrium
is due to derivatives of either $\Lambda$ or $\sigma$.
Section \ref{sec:Fokker-Planck} makes contact with the more common Fokker-Planck
perspective, discussing the relationship of $(\con{u},\con{D})$ with drift and diffusive
currents is discussed, as well as position-conditional mean instantaneous velocity.

\section{Background: Feller processes and semigroups on manifolds}
\label{sec:Feller}

This Section is a quick review of some standard
lore\cite{Gardiner-methods,Kallenberg-foundations}, with two principal
objectives. First, to introduce the semigroup formulation which will be
used subsequently. This is a description which focuses on the evolution
of observables rather than a probability density. It stands in relation
to what might loosely be called a Fokker-Planck-type formulation as the
Heisenberg picture in quantum mechanics stands to the Schr\"odinger picture.
Secondly, and concurrently, to recall the equivalence between Markov processes
with continuous sample paths and drift-diffusions.

For a Markov process $X$ on a manifold ${\cal M}$, and
with $X_t$ denoting the state of the system (a point in ${\cal M}$) at time $t$,
operators $P^t$ are defined via
\begin{equation}
(P^t f)(p) = \Expct_p[ f(X_{t}) ] = \expct[p]{f(t)},
\end{equation}
this being the expectation of $f(X_t)$ assuming initial state $p$.
The markovian no-memory property is expressed by the
Chapman-Kolmogorov equation $P^s P^t = P^{t+s}$ ($s,t \ge 0$),
which can also be read as saying that $(P^t)_{t\in[0,\infty)}$ comprises a semigroup.
A {\it Feller semigroup} is one which maps $C_0({\cal M})$
(continuous functions tending to zero at infinity) into itself
and is strongly continuous with respect to supremum norm, and
its ({\it infinitesimal}) {\it generator} $\Gen$ is defined by
\hbox{$\Gen f \defeq \lim_{t\downarrow 0} \frac{1}{t} (P^t f - f)$,}
with domain the subspace of $C_0({\cal M})$ on which the limit exists.
The generator and semigroup are related by $\frac{d}{dt} P^t f = \Gen P^t f$,
solution to which is {\em formally} written as $P^t = \exp t\Gen$.
Assuming as well that $C_c^\infty({\cal M})\subset \dom \Gen$ 
(so far, all is all consistent with a jump process) and that
all sample paths are continuous,
$\Gen$ is guaranteed to be a second-order differential operator.
In a coordinate chart $\Arr{x}{{\cal U}}{\Reals^n}$, 
\begin{equation}
  \label{eq:Gen}
  \Gen =
  u^i(x) \frac{\partial}{\partial x^i} 
  +  D^{ij}(x) \frac{\partial^2}{\partial x^i \partial x^j}, 
\end{equation}
where
\begin{align}
  \label{eq:u-D}
    u^i(x_0)  & \defeq \Gen [ (x^i-x_0^i) \chi ]
  \nonumber \\
    2D^{ij}(x_0) & \defeq \Gen [(x^i-x_0^i) (x^j-x_0^j) \chi],
  \end{align}
  and $\chi \in C_c^\infty({\cal U})$ equals one on some neighborhood of $x_0$.
  Here and throughout, summation convention is followed for matching
  upper and lower indices.
  Eqs. (\ref{eq:u-D}) provide the interpretation of $(u,D)$.
Exactness of the second-order form goes back 
to a 1931 paper of Kolmogorov\cite{Kolmogorov31}.
Now, while $u^i\partial_i$ is covariant, $D^{ij}\partial_i\partial_j$ is not.
This has the slightly confusing consequence that $D^{ij}$ transforms as a
twice-contravariant tensor,
but the drift has an anomalous term in its coordinate-transformation rule:
\begin{equation}
\label{eq:transformation-law}
u^\alpha(y) =  \frac{\partial y^\alpha}{\partial x^j} u^j(x) +     
\frac{\partial^2 y^\alpha}{\partial x^j \partial x^k} D^{jk}(x).
  \end{equation}
  This curious fact indicates that $(u,D)$ should be thought of as a unit,
  not two fully distinct things.

In the converse direction to the conclusion represented by (\ref{eq:Gen}),
suppose given a differential operator as generator of a semigroup $(P^t)$.
{\it If} $P^t$ bears interpretation as the semigroup associated with a
stochastic process on ${\cal M}$ --- a probability measure or coherent
system of starting-point-indexed probability measures on paths
$[0,\infty)\rightarrow {\cal M}$ --- then the process is perforce markovian
(Chapman-Kolmogorov equation) and has continuous sample paths since
$\Gen f(x)=0$ whenever $f$ vanishes on any neighborhood of $x$.
Further, as already seen, $\Gen$ must in that case be second order.
  
The semigroup formulation is convenient for calculating multi-time correlation functions.
For example, with initial measure $\mu$ and $t_1, t_2 \ge 0$,
\begin{align}
  \Expct_\mu[ f_2(X_{t_1+t_2})  f_1(X_{t_1}) ]
  &= \expct[\mu]{f_1(t_1) f_2(t_1+t_2)}
    \nonumber \\
  & = \int \mu(dx) P^{t_1} f_1 P^{t_2}  f_2(x).
\end{align}
In the final expression, time ordering must be observed, and $P^{t_1}$ acts
on everything to its right (i.e., $f_1 P^{t_2}f_2$).

A general Markov process on ${\cal M}$ can be thought of as a continuous motion
occasionally interupted by unanticipated jumps. Later, we will discuss adding jumps
to our prototype models, but will do no concrete calculations, hence do not introduce
any notation for them.

\section{Contraction of velocity models}
\label{sec:contraction}

This section constructs the contraction of a general velocity model, with
special attention to the markovian limit.
The generator $\con{\Gen}$, equivalently the drift-diffusion field
$(\con{u},\con{D})$ of the contracted process, are expressed in terms
of time correlation functions in the fast equilibrium.
As concrete examples to keep in mind, simple prototypes will be described first.

\subsection{Prototypes}
\label{sec:prototypes}

We introduce some very simple concrete models as examples.
They live on product spaces $\Reals_x^d \times F$, where the first factor is
for position and the second for some internal, fast degrees of freedom.

\noindent $A_0$: $F = \Reals_v^d$ is velocity space and the generator is
\begin{equation}
  \Gen_{A_0}  -\gamma(x) v^i \frac{\partial}{\partial v^i}
  + {D(x)} \Delta_v
+ v^i\frac{\partial}{\partial x^i},
\end{equation}
where $\Delta$ is the $d$-dimensional euclidean laplacian. This is nothing but the
Langevin-Kramers model for a particle in a thermal bath, except that there is
no assumption that anything is thermal, so the fluctuation-dissipation link is
broken and $\gamma$ and $D$ are functions of $x$ (as emphasized in the notation).
The drift in Model $A_0$ was investigated already decades
ago\cite{Ermak+McCammon-78,van-Kampen-88}.

\noindent $B_0$: $F = {\Bbb S}^{d-1}$, the unit sphere
in $\Reals^d$, so a point $\xi$ in $F$ can be thought of as a unit vector in
$\Reals^d$.
$B_0$ has infinitesimal generator 
\begin{equation}
\Gen_{B_0} = {D(x)} \Delta_{\Bbb S} + v_0(x) \xi^i\frac{\partial}{\partial x^i}, 
\end{equation}
where $\Delta_{\Bbb S}$ is the Laplace-Beltrami operator on the sphere.
As motivation, consider that this can model a self-powered colloidal
particle which propels itself by consuming fuel in its environment.
The concentration of fuel, hence its speed $v_0$ varies with the
position-dependent concentration. At the same time, its direction of
motion $\xi$ is subject to Brownian wandering represented by the first
term of $\Gen$.
Variants of this model were investigated relatively
recently\cite{Cates+Tailleur-13,Vuijk+19}.

In Section \ref{sec:elaboration} these models will be significantly
elaborated. But the simple examples are a good thing to keep in mind
during the following abstract development.

\subsection{Slow and fast degrees of freedom}
\label{sec:bundles}

The general and natural geometrical setting for our investigation is as follows.
${Q}$ is a manifold, the configuration, or base, manifold,
representing slow degrees of freedom of the system ---
position in the concrete models.
$\Arr{\pi_E}{E}{Q}$ is a fiber bundle over $Q$ with typical fiber
$F$. $F$ represents fast degrees of freedom. Conventionally, coordinates or points
in the fibers will be denoted by $\xi$.
In the case of models $A_0$ and $B_0$, $E$ is just a product bundle, with
$Q = \Reals_x^d$. Several motivations can be adduced for working in the
manifold setting. Certainly there are interesting cases of particles moving in
noneuclidean geometries, in a lipid bilayer membrane, for example.
Fast variables, such as orientational variables, are perhaps even more likely to
inhabit noneuclidean spaces. Model $B$ of this paper is an example.
Also, since the questions at issue are local in position,
the mathematical overhead involved is fairly light, while nonuniform
parameters are very natural in arbitary coordinates.

What determines the rate of change of position is the velocity, of course.
$\xi$ may simply {\em be} velocity, as for Model $A_0$.
In any case, the velocity is a function $v(x,\xi)$ of the fast
and slow coordinates. Thus, $v\colon E \rightarrow TQ$ is a bundle map.
The basic assumption made about the fundamental Markov process is that it
have a generator of the form
\begin{equation}
\label{eq:generator-split}
\Gen = \Genint + v^i(\xi,x) \frac{\partial}{\partial x^i},
\end{equation}
where $\Genint$ generates a fiber-preserving semigroup $P_{\text{fst}}^t$
(the state never moves out of the fiber it started in) in each fiber.
The structure of (\ref{eq:generator-split}) explains the name ``velocity model''.
In addition, the fast processes are assumed to attain equilbrium on the time
scale $\tau$:
\begin{equation}
  \label{eq:internal-equilibrium}
  P_{\text{fst}}^t f = \efst{f} + \bigO{e^{-t/\tau}}.
\end{equation}
That is, $\efst{\cdot}$ represents a fiber-by-fiber equilibrium 
depending on base position (in $Q$).
%

The smaller $\tau$ compared to our observation time scale
for the contracted process, the closer the latter is to markovian.
In order to access the markovian limit in a formal
way, we scale $\Genint$ up by a large factor, scaling down the time required
to reach fast equilibrium. That alone would have the effect of rendering
all fluctuations negligible, so a rescaling of the velocity is also needed.
With temporary ``old/new'' labels for clarity, we take
\begin{align}
&   \Gen_{\text{fst,new}} \defeq \frac{1}{\epsilon}\Gen_{\text{fst,old}},
  \nonumber \\
  &  v_{\text{new}}(\xi,x) - \efst{v_{\text{old}}} \defeq
    \epsilon^{1/2} (v_{\text{old}}(\xi,x) - \efst{v_{\text{old}}})
  \label{eq:rescaled}
\end{align}
For infinitesimal $\epsilon$, the {\it contracted} process is
markovian, and therefore, according to Section \ref{sec:Feller},
fully characterised by drift-diffusion fields $(\con{u},\con{D})$ on $Q$.
An overbar on a symbol indicates that it pertains to a contracted process.

\subsection{Contracted semigroup}
\label{sec:perturbation}

The fast generator $\Genint$ generates a fast semigroup $\Pfst$ and drives
the system toward the fast equilibrium $\efst{\;}$. These are all black boxes
to us. In such generality, there are very few options of how to proceed.
Take the clutching term
\begin{equation}
  \delta\Gen = v\cdot{\partial_x} \defeq v^i(x,\xi)\frac{\partial}{\partial x^i},
\end{equation}
from $\Gen$, 
and put it into the Duhamel-Dyson formula as a perturbation:
\begin{align}
\label{eq:Duhamel-Dyson}
P^T
&  = \Pfst^T + \int_0^T P^{s} (\delta\Gen) \Pfst^{T-s} \, ds
 \\
 & =  \Pfst^T
         + \int_{0 \le t_1 \le T} \Pfst^{t_1} (\delta\Gen) \Pfst^{T-t_1}
         \nonumber
          \\
       & + \int_{0\le t_2\le t_1 \le T} \Pfst^{t_2} (\delta\Gen) \Pfst^{t_1-t_2} (\delta\Gen) \Pfst^{T-t_1}
         + \cdots
         \nonumber
\end{align}
So far, this resembles perturbation expansions for evolution operators as they
occur in many fields, quantum mechanics notably.
However, we wish to apply this to functions of position (in $Q$) {\em alone},
thereby constructing a contracted semigroup $\con{P}^t$ with generator $\con{\Gen}$.
To that end, examine the structure of the $n$\textsuperscript{th} term of the expansion:
\begin{align}
\int_{0\le t_n\le \cdots \le t_1 \le T}
  & \Pfst^{t_n} (\delta\Gen) \Pfst^{t_{n-1}-t_n} (\delta\Gen) \cdots
    \nonumber \\
& \cdots \Pfst^{t_2-t_3} (\delta\Gen) \Pfst^{t_{1}-t_2} (\delta\Gen) \Pfst^{T-t_1}.
\end{align}
The integrand is a string of operators, $\Pfst^s$ and $(\delta\Gen)$,
each of which potentially acts on everything to its right.
If $\epsilon\tau \ll {t_k-t_{k-1}}$, that action is effectively the same as
taking the fast-equilibrium value. Thus, the integrand reduces to a product of
clusters of the form
$\efst{ \text{ sequence of } \delta\Gen\text{'s within times } \sim\epsilon\tau}$.
The entire perturbation expansion thus reduces, formally, to a sum over all
(ordered) products of clusters as the cluster positions range over the interval
$[0,T]$. That is,
\hbox{$\con{P}^T = \exp(T\con{\Gen})$,} with
\begin{align}
  \label{eq:enchilada}
\con{\Gen} &= \sum_{n=1}^\infty \efst{S_n},
 \\
S_1 &= v\cdot{\partial_x},
                 \nonumber \\
  S_{n+1} &= ({v\cdot{\partial_x}})\int_0^\infty P^t \Big( S_n - \efst{S_n} \Big).
            \nonumber
\end{align}
Insofar as we are interested in the markovian limit, however,
this is overkill. Each factor of $\delta\Gen$ carries a factor $\epsilon^{-1/2}$
in the $v$, and each integration $\int_0^\infty \Pfst^t\cdots dt$ brings another factor
$\epsilon$. Formally, then, $\efst{S_1} = \efst{v}\cdot{\partial_x} = \bigO{1}$,
and, for $n\ge 0$, $\efst{S_{n+2}} = \bigO{\epsilon^{n/2}}$.
Therefore, in the limit $\epsilon \downarrow 0$, we obtain the markovian generator
\begin{align}
  \label{eq:contracted-markovian-gen}
  \con{\Gen}_{\text{markovian}} = & \efst{v}\cdot{\partial_x}
  \\
     & + \int_0^\infty dt\, \efst{  \left( \tilde{v}\cdot{\partial_x} \right)
                                \Pfst^t \tilde{v}\cdot{\partial_x} },
                            \nonumber
\end{align}
using the abbreviation
\begin{equation}
  \nonumber
\tilde{v} \defeq v - \efst{v}.
\end{equation}
(The subtraction of the first $\efst{v}$ on the second line is allowed since what
it multiplies has zero fast expectation value.)
Henceforth, we are interested only in this markovian limit and therefore
drop the subscript.

Now, recalling once again that the derivatives in (\ref{eq:contracted-markovian-gen})
act on everything to the right and matching against the general form (\ref{eq:Gen}),
the drift-diffusion fields of the contracted process in the markovian limit
are revealed:
\begin{equation}
  \label{eq:con-u}
\con{u} = \efst{v} +   
\int_0^\infty dt\, \efst{ \tilde{v}(0)\cdot({\partial_x} \tilde{v}(t)) },
\end{equation}
and
\begin{equation}
  \label{eq:con-D}
\con{D} = 
2 \int_0^\infty dt\, \Sym \efst{ \tilde{v}(0) \otimes \tilde{v}(t) },
\end{equation}
where `Sym' indicates symmetrization on the tensor indices.
Verification that these formulas conform to the coordinate-transformation law
(\ref{eq:transformation-law}) is straightforward.

\section{Prototypes elaborated}
\label{sec:elaboration}

This Section explores various ways to elaborate the prototype models $A_0$ and $B_0$.
Rather than doing so in an unconstrained fashion, the aim is to preserve a simplicity
of the relevant correlation functions.
Thus, in preparation we define a restricted class of velocity models with
a convenient and fairly intuitive set of parameters:
intrinsic drift ($\efst{v}$), decay ($\Lambda$), velocity transformation ($\sigma$),
and kinetic ($K$) tensors. Definitions follow.

\subsection{Quasilinear models}
\label{sec:quasilinear}

quasilinear models are defined by two conditions.
\newline
(a) The velocity function has a factorized form:
\begin{equation}
  \label{eq:factorized}
\tilde{v}^i(\xi,x) = {\sigma^i}_j(x) \tilde{v}_0^j(\xi);  
\end{equation}
(b) $\tilde{v}$ decays exponentially, on average:
\begin{equation}
\label{eq:decay}
\Genint\tilde{v}^i = -{\Lambda^i}_j \tilde{v}^j,
\end{equation}
where $\Lambda$ is a position-dependent linear operator with spectrum in the right complex
half-plane $\mathrm{Re}\, z \gtrsim \tau^{-1}$.

Condition (b) implies that
\begin{equation}
\nonumber
\Pfst^t \tilde{v} = e^{-t\Lambda} \tilde{v}.
\end{equation}
The calculations for $\con{u}$ and $\con{D}$ under these assumptions are similar,
the latter being slightly easier.
For $\con{u}$ (\ref{eq:con-u}), we have
\begin{align}
\con{u} - \efst{v} =  & 
     \int_0^\infty \efst{ \tilde{v}(0)\cdot\partial_x  \tilde{v}(t) }
   \nonumber \\
  &=
 \int_0^\infty \efst{ (\tilde{v}(0)\cdot\partial_x)  \exp\left(-t{\Lambda}\right)\tilde{v} }
  \\
  \label{eq:con-u-quasilinear-0}
& = \efst{ (\tilde{v}\cdot\partial_x)  {\Lambda}^{-1} \tilde{v} }.
\end{align}
Now, according to the factorization condition (a), $\partial_x$ here is oblivious
to $\tilde{v}_0$, so we can rearrange factors and put it before the derivative.
The result will be correct in any specific coordinate system.
({\it Vide infra} for comment about transformation properties.)
With the definition
\begin{equation}
  \label{eq:K-def}
  K \defeq \efst{ \tilde{v} \otimes \tilde{v} },
\end{equation}
we obtain
\begin{align}
  \con{u}^i - \efst{v^i}
  &= {(\sigma^{-1})^k}_s K^{sj} \frac{\partial}{\partial x^j}
    \left[ {(\Lambda^{-1})^i}_l {\sigma^l}_k\right],
  \label{eq:con-u-quasilinear}
   \\
2\con{D}^{ij} 
  &=  {(\Lambda^{-1})^i}_k K^{kj} +  {(\Lambda^{-1})^j}_k K^{ik}.
  \label{eq:con-D-quasilinear} 
\end{align}

The expression for the generator corresponding to this $(\con{u},\con{D})$
pair is actually hardly less compact than either. 
It is
\begin{equation}
  \label{eq:quasilinear-con-Gen}
  \con{\Gen} = \left(
    \efst{v^i}
    + (\sigma^{-1} K)^{kj} \,\partial_j\, {(\Lambda^{-1}\sigma)^i}_k  
  \right) \partial_i.
\end{equation}
Here, the alternative notation for derivatives is meant to emphasize that
$\partial_j$ is an operator that operates on everything to its right, not
just the parenthesized expression immediately following.
Written in terms of fast-equilibrium correlations, this is
\begin{equation}
  \label{eq:quasilinear-con-Gen-abstract}
  \con{\Gen} = 
    \efst{v^i}\partial_i
    + \efst{ \tilde{v}^j \partial_j {(\Lambda^{-1})^i}_k \tilde{v}^k }\partial_i.
  \end{equation}
  This expression is valid with only condition (b) imposed, but that is not the
  reason it is written here. (\ref{eq:quasilinear-con-Gen-abstract}) exhibits
  general coordinate covariance, whereas (\ref{eq:quasilinear-con-Gen}) does not.
  The reason is that, in the latter, a factor $\tilde{v}_0$ was moved in front
  of a derivative so that we could write things in terms of $K$.
  $\tilde{v}_0$ itself is not position-dependent, but with a change of
  coordinates, it will generate a jacobian factor which may be, but which
  is invisible in (\ref{eq:quasilinear-con-Gen}).

  Before moving on to concrete models, a few observations are in order about the
  significance of (\ref{eq:con-u-quasilinear}) and (\ref{eq:con-D-quasilinear}).
(i) Apart from $\efst{v}$, there are two basic mechanisms to produce drift in
  a quasilinear model:
gradients of either the decay tensor $\Lambda$ or the velocity scaling tensor $\sigma$.
These two are precisely the mechanisms whereby drift is generated in
the prototypes $A_0$ and $B_0$, respectively.
(ii) In contrast to the case of a generic velocity mdoel, the quasilinear
parametrization allows a precise statement of that sort.
Still, it seems reasonable to say that something roughly like that holds in general.
(iii) Derivatives of $K$ do not enter. Therefore, we cannot say that
inhomogeneity generically induces drift. This is not in conflict with the
existence of thermophoresis\cite{Piazza+Parola-08}.
Temperature-dependent free energy of interaction
with the surrounding medium can contribute to $\efst{v}$.
(iv) The only information required directly about the fast equilibria is
that encoded in $\efst{v}$ and $K$, which is to say, not very much.

  \subsection{Model $A$}
\label{sec:model-A}

Now, we briefly investigate what sorts of physically plausible elaborations can be
made of Model $A_0$ (later, $B_0$) from Section \ref{sec:prototypes} 
without spoiling the quasilinearity conditions (\ref{eq:factorized}) and (\ref{eq:decay}).
The first, most obvious way to modify Model $A_0$ is simply to make
all the scalar coefficients tensors. Thus, Model A is defined by the
fast generator ($\gamma$, $f$, and $D$ are functions of $x$ alone)
\begin{equation}
  \label{eq:model-A-gen}
  \Gen_{A,\text{fst}} = (f^j-{\gamma^j}_{i} v^i) \frac{\partial}{\partial v^j}
  + D^{ij}\frac{\partial^2}{\partial v^i\partial v^j}.
\end{equation}
Here, $f$ has the interpretation of some kind of external force.
$\gamma$ is allowed to be anisotropic and may have partially rotational character,
and $D$ is symmetric and positive, but otherwise unconstrained.
With regard to the anisotropies, this is not written down with any specific physical
system in mind, but one may imagine motion in an anisotropic medium, e.g. nematic
liquid crystal, a magnetic field, and random excitation coming from an anisotropic
nonthermal source such as ultrasonic pulses.

Three of the quasilinear parameters follow immediately:
\begin{equation}
  \label{eq:model-A-3-params}
  \sigma\equiv I,
  \quad \efst{v} = \gamma^{-1} f,
  \quad \Lambda = \gamma.
\end{equation}
 Finally, demanding $\efst{\Gen_A (\tilde{v}\otimes\tilde{v})} = 0$ gives the equation
\begin{equation}
  \label{eq:model-A-K-implicit}
\Lambda K + K \transpose{\Lambda} = 2D,  
\end{equation}
which determines $K$ implicitly, and has a unique solution given the
spectral restrictions on $\Lambda$.
In fact, writing $K$ and $D$ as bilinear forms,
(\ref{eq:model-A-K-implicit}) says $K({\Lambda} v,w)
+ K(v,{\Lambda}w) = 2D(v,w)$.
The solution is $K(v,w) = \int_0^\infty 2D( e^{-t{\Lambda}}v, e^{-t{\Lambda}}w)\, dt$,
which is manifestly symmetric and positive, since $D$ is.

Multiplying (\ref{eq:model-A-K-implicit}) by $\Lambda^{-1}$ and
$\transpose{\Lambda}^{-1}$ from left and right, respectively, yields
\begin{equation}
\label{con-D-model-A}
\con{D}^{ij} =  {(\Lambda^{-1})^i}_k  {(\Lambda^{-1})^j}_m  D^{km}, 
 \end{equation}
bypassing $K$.
Unfortunately, $K$ is still needed for $\con{u}$,
and no tidy formula appears available in general.
Three special cases will be considered.

Model $A_1$: $\gamma = \gamma_s I$ is proportional to the identity 
(subscript `$s$' for `scalar').
Then, the foregoing quickly leads to
\begin{equation}
  \label{eq:A1-con-Gen}
  \con{\Gen}_{A_1}
  = \Big( \gamma_s^{-1} f^i +
  \gamma_s^{-1} D^{ij} \,\partial_j\, \gamma_s^{-1}
  \Big) \partial_i
\end{equation}
We might try to further specialize by supposing $D$ is scalar.
That makes no sense in general, but if $Q$ has a riemannian metric $g$,
we may suppose $D^{ij}$ and $K^{ij}$ proportional to $g^{ij}$.
Write $K^{ij} = T g^{ij}$, so $D^{ij} = (T/\gamma_s)g^{ij}$; this can be viewed
as a definition of $T$.
Essentially, this is Model $A_0$ with an external force.
The result here is easily seen to agree with that obtained by
van Kampen in Section 4 of Ref. \onlinecite{van-Kampen-88} after notational
translation.

Model $A_2$: $Q$ is equipped with a riemannian metric $g$ and $K^{ij}=Tg^{ij}$.
The interpretation of this is that fast-equilibrium is thermal equilibrium at
position-dependent temperature $T(x)$, and relies on something outside the
model itself for its justification. 
Thus, the formula for $K$ is given by fiat, and implies the restriction
\begin{equation}
  \label{eq:model-A-thermal}
  D^{ij} = \frac{T}{2}\left[ \gamma^{ij} + \gamma^{ji} \right],
\end{equation}
where one index of $\gamma$ is raised with $g$ in the usual way.
From this, we easily obtain
%
\begin{equation}
  \label{eq:A2-con-Gen}
  \con{\Gen}_{A_2}
  = \Big( (\gamma^{-1} f)^i +
  T g^{kj} \,\partial_j\, {(\gamma^{-1})^i}_k
  \Big) \partial_i
\end{equation}
Model $A_3$: $\Lambda$ has a full complement of smoothly varying position-dependent
(complex!) eigenvectors labelled by $\alpha = 1,\ldots,d$:
\begin{equation}
  \label{eq:case-3-eigvecs}
{\Lambda^i}_k {\theta_\alpha}^k = \lambda_\alpha {\theta_\alpha}^i.  
\end{equation}
With projection operators defined as
\begin{equation}
P_\alpha = {\theta^\alpha}\otimes{\theta_\alpha}
\end{equation}
by use of the dual basis $(\theta^\alpha)_1^d$ to the basis $(\theta_\alpha)_1^d$,
\begin{equation}
{\theta^\alpha}_i {\theta_\beta}^i  = \delta_{\alpha\beta},
\end{equation}
we obtain
\begin{equation}
\label{eq:case-3-Lambda}
{\Lambda} = \sum_{\alpha=1}^d \lambda_\alpha P_\alpha,
\quad
{\Lambda}^{-1} = \sum_{\alpha=1}^d \frac{1}{\lambda_\alpha} P_\alpha.
\end{equation}
Now, we can proceed to write everything else in terms of the eigenvalues
$\lambda_\alpha$ and projections $P_\alpha$. 
The results are not particularly transparent, but from (\ref{eq:model-A-u-case-3})
below, one at least gets some sense of how derivatives of both the eigenvalues and
the eigendirections contribute to the drift.
The kinetic tensor is
\begin{equation}
  \label{eq:model-A-K-case-3}
K^{ij} = \sum_{\alpha,\beta=1}^d
\frac{ 2D^{km} {{P_\alpha}^i}_k  {{P_\beta}^j}_m }{\lambda_\alpha + \lambda_\beta},
\end{equation}
the contracted diffusion tensor is
\begin{equation}
  \label{eq:model-A-K-case-D}
\con{D}^{ij} = 
\sum_{\alpha,\beta=1}^d
\frac{D^{km} {{P_\alpha}^i}_k  {{P_\beta}^j}_m }{(\lambda_\alpha + \lambda_\beta)^2},
\end{equation}
and the contracted drift is
\begin{equation}
  \label{eq:model-A-u-case-3}
\con{u}^{i} = {(\gamma^{-1})^i}_kf^k  
+ \sum_{\alpha,\beta,\gamma=1}^d
\frac{D^{nm} {{P_\alpha}^k}_n  {{P_\beta}^j}_m }{\lambda_\alpha + \lambda_\beta}
\frac{\partial}{\partial x^j}\left(\lambda_\gamma^{-1}{{P_\gamma}^i}_k\right).
\end{equation}

Consider briefly some additional elaborations that would preserve 
quasilinearity without writing formulas for the parameters.
First, $D$ can be made velocity dependent. 
Secondly, a variety of jumps can be included ---
a mean zero jump distribution would contribute to $K$,
while jumps with probability density depending on a scattering angle and 
fractional decrease of $|v|$ would contribute to $\Lambda$.

\subsection{Model $B$}
\label{sec:model-B}

The elaboration of Model $B_0$ is perhaps more interesting.
As for Model $A_2$, we assume here that $Q$ is a riemannian
manifold.
The prototype, $B_0$, was defined on the sphere bundle $Q\times {\mathbb S}^{d-1}$;
$\xi$ was just a unit vector in that case.
Implicitly, the particle was supposed to have only a single distinguished axis,
that along which it propels itself.
The full Model $B$ will be defined on the associated orthonormal frame bundle;
$\xi$ will be an entire orthonormal frame keeping track of all the orientational
degrees of freedom.
Consider a particle shaped like a brick.
This is easy to talk about and visualize,
but an ellipsoid, or even more complicated shape with appropriate
reflection symmetries, will have the same type of rotational characteristics.
A brick has three principal axes for its hydrodynamic properties with, as a result,
distinct rotational diffusion coefficients about each of them.
A dynamically inert dot painted on one of each pair of opposite faces
of the brick allows unambiguous definition of an {\em oriented} axis and hence
the association of a distinct orthonormal frame with each particle configuration.
In the $d$-dimensional case, the body-frame $\alpha\beta$
plane has its own distinct orientational diffusion coefficient $D_{\alpha\beta}$
($D_{\beta\alpha} = D_{\alpha\beta}$, diagonal entries zero).

This picture is intuitively simple.
However, some notation needs to be set up to write
down formulas. Given a reference frame $(\evec{1},\ldots,\evec{d})$, each
body frame can be identified with the rotation matrix $\xi$ in $SO(d)$ which
brings the reference frame into coincidence with it. That is, $\xi\in SO(d)$ is
identified with the frame
$(\xi_{1},\ldots, \xi_{d})$, where $\xi_n$ is the $n$\textsuperscript{th}
column of $\xi$ with components ${\xi^i}_n$.
Now, let $\llbracket \alpha\beta;\theta \rrbracket$ denote the 
rotation by angle $\theta$ in the reference $\alpha\beta$ plane, and define
the operators $\tilde{L}_{\alpha\beta}$ for $\alpha\neq\beta$ via
\begin{equation}
  \label{eq:L_ab}
  (\tilde{L}_{\alpha\beta}f)(\xi)
  \defeq \frac{d}{d\theta}
  f(\xi\llbracket \alpha\beta;\theta\rrbracket)\Big|_{\theta=0}.
\end{equation}
This differentiates with respect to rotation about the {\em local},
configuration-dependent, $\alpha\beta$ plane for each $\xi$.
For computational purposes, it is useful to note that
$\tilde{L}_{\alpha\beta}$ is a derivation: 
$\tilde{L}_{\alpha\beta}(fg) = (\tilde{L}_{\alpha\beta}f)g + f (\tilde{L}_{\alpha\beta}g)$;
and 
\begin{equation}
\label{eq:L_ab-practical}
\tilde{L}_{\alpha\beta}\xi = \xi E_{\alpha\beta},
\end{equation}
where $E_{\alpha\beta}$ is a matrix with $\alpha\beta$ component equal to $1$,
$\beta\alpha$ component equal to $-1$ and all other components zero.
%

Finally, we can write down the generator for fully anisotropic
orientational diffusion:
\begin{equation}
  \label{eq:B-Gen}
\Gen_{B,\text{fst}} = \sum_{\alpha < \beta} D_{\alpha\beta} \tilde{L}_{\alpha\beta}^2.
\end{equation}
To finish the specification of the model, assume that the velocity is along
$\evec{1}$ in the reference configuration, so that its direction in
configuration $\xi$ is ${\xi}_1$, and
the velocity map is
\begin{equation}
  \label{eq:model-B-velocity-map}
v^i(\xi,x) = {\sigma^i}_j(x) {{\xi}^j}_1.
\end{equation}
It may seem that there is no possible physical interpretation for
nonscalar $\sigma$, but it will be put to good use in Section \ref{sec:two-stage}.

For Model $B$, the fast process induced on the tangent space $T_xQ$ via
the velocity map $v(\cdot,x)$ is not generally markovian.
To see that, suppose that $D_{1,2}$ is much
larger than all other orientational diffusion coefficients.
And, suppose that we have observations of $v$ at two times with a delay
of order $D_{1,2}^{-1}$. Then, we have a good idea of where the local
$2$-axis is and can expect the velocity to keep moving approximately in 
the $1$-$2$ plane for the near future.
Nevertheless, the induced process is quasilinear, as follows from a
straightforward calculation
yieldng
\begin{equation}
  \nonumber
\Gen_{B,\text{fst}}\, {\xi}_1 = - (d-1)D_{\text{rot}} {\xi}_1,
\end{equation}
with the definition
\begin{equation}
  \label{eq:model-B-D}
D_{\text{rot}} \defeq \frac{1}{d-1}\sum_{1 < \beta} D_{1\beta}.
\end{equation}

In case $D_{1\beta} = D_{\text{rot}}$ for $2\le \beta \le d$,
the motion of the orientation $\xi_{1}$ is ergodic,
other rotational degrees of freedom are decoupled, and the
model effectively reduces to the prototype $B_0$.

Along with $\sigma(x)$, given {\it a priori}, the quasilinear parameters are
then
\begin{align}
  \label{eq:model-B-params}
{\Lambda^i}_j(x)  &= (d-1)D_{\text{rot}} \, {\delta^i}_j
                  \nonumber \\
 K^{ij}(x)
         & = \frac{1}{d}  {\sigma^i}_k{\sigma}^{jk}.
\end{align}
Substitution into the quasilinear formula (\ref{eq:quasilinear-con-Gen}) yields
\begin{equation}
  \label{eq:B-con-Gen}
 \con{\Gen}_B
  = \frac{1}{d(d-1)} 
    {\sigma^j}_k \,{\partial_j}\, \frac{ {\sigma^i}_k}{D_{\text{rot}}} \partial_i.
\end{equation}
The formula for $K$ follows from 
$\efst{{\xi^j}_i {\xi^k}_i} = \delta_{jk}/d$, or more strongly from
equidistribution of $\xi_1$ over ${\mathbb S}^{d-1}$
in fast equilibrium, which the reader may consider evident enough.
For a more formal derivation,
consider $0 = \efst{\Gen_{B,\text{fst}}{\xi^j}_i {\xi^k}_i} = 
\sum_n D_{in}[\efst{{\xi^j}_n {\xi^k}_n}  - \efst{{\xi^j}_i {\xi^k}_i}]$.
This implies that
$\efst{{\xi^j}_i {\xi^k}_i} = (\sum_n D_{in})^{-1} \sum_n D_{in}\efst{{\xi^j}_n {\xi^k}_n}$
is independent of $i$ (each of a set of expectations is a weighted average of the others).
Two appeals to orthogonality of $\xi$ after summing, first over $i$ and then over $j=k$,
show that $\efst{{\xi^j}_i {\xi^k}_i} = \delta_{jk}/d$.

Due to the interest for run-and-tumble particles\cite{Cates+Tailleur-13},
addition of jumps to the model are now briefly considered.
In keeping with the operative symmetries, the jump rate measure $\nu(\xi; d\eta)$
(rate of transitions $\xi \mapsto d\eta$)
is required to have a well-defined form with respect to the body frame,
and to be invariant under reflection $r_i$ through the hyperplane normal to
the body axes $\xi_{i}$, $i=2,\ldots,d$:
\begin{equation}
\nu(\xi;\xi\, d\eta) = \nu_0(d\eta),  
\quad \nu_0(r_i\, d\eta \, r_i)  = \nu_0(d\eta).
\end{equation}
With $|\nu| = \int \nu_0(d\eta)$ denoting the total jump rate,
the associated contribution to the generator is defined via
\begin{equation}
\Gen_{\text{jmp}} f(\xi) = -|\nu| f(\xi) + \int f(\xi\eta) \nu_0(d\eta).  
\end{equation}
In particular, using reflection symmetry in the second equality,
\begin{equation}
  \Gen_{\text{jmp}} \evec{1}
  =  -|\nu|\evec{1}
    + \int \eta\evec{1} \nu_0(d\eta) = -\alpha |\nu| \evec{1},
  \end{equation}
  where $0 \le \alpha \le 2$.
  For scattering which sends $\evec{1}$ uniformly over the sphere
  or pure reversal, $\alpha =$  $1$ or $2$, respectively.
  In the markovian limit, therefore, jumps simply make an additive
  contribution to the effective diffusion coefficient (and thence to $\Lambda$).
  $D_{\text{rot}}$ is already an effective quantity insofar as the various $D_{1\beta}$
  are not necessarily equal. With jumps, it should be replaced by
\begin{equation}
\nonumber
D_{\text{rot}} + \frac{\alpha |\nu|}{d-1}.
\end{equation}

\subsection{A two-stage contraction}
\label{sec:two-stage}

Now, consider combining two velocity models, 1 and 2.
A trivial way to do this is to define a (sum) velocity model via
$v(\xi_1,\xi_2,x) = v_1(\xi_1,x) + v_2(\xi_2,x)$, where the fast parts of
models 1 and 2 are independent. Then the sum model is not quasilinear,
but that is not a problem, because each of models 1 and 2 makes its
own separate contribution to $\bar{u}$ and $\bar{D}$
(i.e., we should add $\bar{u}$ and $\bar{D}$).

If models 1 and 2 have very different relaxation times, however,
a more interesting possibility --- two-stage contraction --- emerges.
Let us consider an example constructed from Models $A_2$ and $B$.
It will be essentially the same model as treated by Vuijk {\it et alia}\cite{Vuijk+19},
except that we allow inhomogeneous $T$ and $D_{\text{rot}}$.
Following that source, the model is motivated as follows.
Consider an active colloidal particle ($\sim 1 \mu$m)
a self-generated {\em thrust} $f_0$ in the direction ${\xi}_1$
moving in a liquid.
Ordinarily, the velocity would simply be some scalar multiple of that,
in fact $ (f_0/m \gamma) {\xi}_1$, where $\gamma^{-1}$ is the viscous
relaxation time for velocity, which, for a micron-sized particle in
aqueous medium, is of order 1 $\mu$s.
Really, we should understand that velocity as the output of a
Langevin-Kramers model with the external driving force being
the particle's self-generated thrust. But, what if the damping tensor
$\gamma$ is not a simple scalar? The scenario envisaged by
Vuijk {\it et alia} ({\it loc. cit.}) is a charged particle in a magnetic field, but
we do not commit to that and allow general $\gamma$.

Since the relaxation time of the Model $A_2$ velocity in this scenario is
of order $10^{-6}$ times that of the thrust direction, we first eliminate the
Model $A_2$ variables. 
According to Eq. (\ref{eq:A2-con-Gen}) ($m=1$), that gives
\begin{equation}
  \nonumber
  \con{\Gen}_{A}
  =  f_0 (\gamma^{-1} {\xi}_1)^i \partial_i
  + T g^{kj} \,\partial_j\, {(\gamma^{-1})^i}_k\partial_i.
\end{equation}
The second term does not depend on $\xi$ so we set it aside temporarily.
The coefficient in the first term is now taken as the velocity function of Model $B$:
$v^i(x,{\xi}_1) = f_0 (\gamma^{-1} {\xi}_1)^i$, equivalently,
$\sigma = f_0 \gamma^{-1}$, where {\em both} $f_0$ and $\gamma$ are position-dependent.
Inserting this data into (\ref{eq:B-con-Gen}) immediately yields
\begin{equation}
  \nonumber
 \con{\Gen}_B
  = \frac{f_0}{d(d-1)} 
    {(\gamma^{-1})^j}_k \,{\partial_j}\, \frac{f_0 {(\gamma^{-1})^i}_k}{D_{\text{rot}}} \partial_i.
  \end{equation}
  Adding this to the set-aside piece of $\con{\Gen}_A$ gives the fully contracted
  generator
\begin{equation}
  \label{eq:2-stage-con-Gen}
 \con{\Gen}
 =
\Big( \frac{f_0}{d(d-1)} 
    {(\gamma^{-1})^j}_k \,{\partial_j}\, \frac{ f_0{(\gamma^{-1})^i}_k}{D_{\text{rot}}} 
 + T g^{kj} \,\partial_j\, {(\gamma^{-1})^i}_k \Big) \partial_i.
  \end{equation}
Note that all parameters here ($f_0$, $\gamma$, $T$, $D_{\text{rot}}$, and $g$)
are potentially position-dependent.
In the case of 3-dimensional euclidean space with uniform $T$, $D_{\text{rot}}$, and $f_0$,
this agrees with Eq. (12) of Ref. \onlinecite{Vuijk+19}, where it was obtained
by very different methods.
(Note: ${\gamma^i}_j$ here corresponds to $\gamma {\Gamma^i}_j$ of Vuijk {\it et alia}.)

\section{Contacting the Fokker-Planck perspective}
\label{sec:Fokker-Planck}

This section makes contact with the Fokker-Planck perspective.
Section \ref{sec:currents} takes a look at the current and asks what, if
anything, the approach taken here says about splitting it into
drift and diffusion parts. Section \ref{sec:instantaneous}
returns to the methods of Section \ref{sec:contraction} to derive
the conditional (on position) mean instantaneous velocity.
Throughout, $Q$ is assumed to be equipped with a riemannian metric $g$.
That allows the definition of a scalar probability density $\rho$ with
respect to volume measure $d\Omega = \sqrt{|g|} dx^1\cdots dx^n$.
Thus, $\rho$ is independent of coordinates.

\subsection{Fokker-Planck equation and currents}
\label{sec:currents}

This subsection considers a drift-diffusion which may not have come from
contraction, so the overbars on $(u,D)$ have been dropped.

Via integration by parts, the time evolution of the expectation of test
function $f(x)$ is found to be
\begin{align}
  \label{eq:d<f>/dt-integrated}
  \frac{d}{dt} \expct[\rho]{f}
  &=  \int \rho \Big( {u}^i\partial_i  + {D}^{ij}\partial_i\partial_j\Big) f\, d\Omega
                          \nonumber \\
  &=  \int f (-\DIV J) \, d\Omega.
\end{align}
This is simply an implicit way of writing the
the Fokker-Planck equation $\frac{\partial \rho}{\partial t} = -\DIV J$.
Here,
\begin{equation}
  \label{eq:current-density}
  J^i[\rho] \defeq
  {u}^i\rho  - \DIV (\evec{i}\cdot{D}\rho), 
\end{equation}
is the probability current density
with $\evec{i}$ the unit vector field along the $i$\textsuperscript{th} coordinate
direction and 
$\DIV V = {|g|}^{-\frac{1}{2}} \partial_i({|g|}^{\frac{1}{2}}V^i)$ denotes the
covariant divergence of the contravariant vector field $V$.
A possibly more aesthetically pleasing way to express the current density is
implicitly via
\begin{equation}
  \int \eta_i J^i \, d\Omega
  = \int \Big\{ [u^i+\Gamma^i_{kj}D^{kj}]\eta_i + D^{ij}\nabla_j\eta_i 
  \Big\} \rho d\Omega,
\end{equation}
where $\nabla_j$ denotes a covariant derivative and $\Gamma^i_{jk}$ is
a Christoffel symbol. Since the anomalous terms in the transformations of
$u$ and $\Gamma$ cancel, the object in square brackets is a genuine covariant vector field.

The expression for the current is trivially rearranged as
\begin{equation}
  \label{eq:rearranged-current}
  J^i[\rho] =
  \left({u}^i - \frac{\partial {D}^{ij}}{\partial x^j}\right)\rho
  - {D}^{ij}
  {|g|}^{-\frac{1}{2}} \frac{\partial}{\partial x^j} ({|g|}^{\frac{1}{2}} \rho).
\end{equation}
The literature seems to show a preference for rearranging the right side of
(\ref{eq:current-density}) as (\ref{eq:rearranged-current}) and identifying the first term as
the total drift, with $-{\partial {D}^{ij}}/{\partial x^j}$ being an
``extra drift''\cite{van-Kampen-88}. I argue now against that identification,
recognizing that, insofar as only the total current is physically relevant, it is a
relatively low-stakes issue. Let us move to euclidean space to keep matters simple,
and visualize the density $\rho$ as an ensemble of noninteracting particles.
``Drift'' ought to refer to a propensity of any one of these particles to move
in a particular direction.
According to the defintion (\ref{eq:u-D}), $u \, dt$ provides us with an unbiased
estimate of where a randomly selected member of the ensemble will go, while $D \, dt$
quantifies the uncertainty. Furthermore, there is nothing strange about a
{\em diffusive} current arising from a gradient of $D$ with uniform $\rho$.
Consider the familiar, elementary explanation of ordinary diffusive current:
a plane $P$ separates a low-density region on the left from a high-density region
on the right. Particles move randomly away from wherever they are.
More particles randomly move across from right-to-left than vice versa, hence
a diffusive current. If now we have equal densities, but more vigorous random
motion on the right (larger $D$), a right-to-left current will again arise.
Thus, it is perfectly proper to consider $(\nabla D) \rho$ a contribution to
{\em diffusive} current.
All this may have a whiff of ``It\^o versus Stratonovich''\cite{van-Kampen-81}
about it. However, since the approach taken here has neither the need for
an interpretational rule that a Langevin equation does, not the complication
of having a general probability distribution from the beginning, it is difficult
to see any interpretational wiggle room.

\subsection{Conditional mean instantaneous velocity}
\label{sec:instantaneous}

Now suppose a velocity model with contracted probability density
$\rho(x)$ which is well-aged (i.e., has been freely evolving for
many fast-relaxation times $\tau$). What is the joint distribution of position and velocity? 
Refs. \onlinecite{Vuijk+19,Cates+Tailleur-13} used a direction calculation of
that joint distribution to find Fokker-Planck equations for Model $B$ variants.
The method of Section \ref{sec:contraction} avoided it, but it is nevertheless
a question of independent interest.
One's first thought might be that the joint distribution takes the
factorized form
\begin{equation}
  \label{eq:factorized-density}
  (\rho\ltimes\rho_{\text{fst}})(x,v)  \defeq \rho(x)\rho_{\text{fst}}^x(v) 
\end{equation}
where $\rho_{\text{fst}}^x$ denotes the fast-equilibrium distribution at $x$.
Actually, the distribution over $v$ is slightly skewed from this,
even in the markovian limit.
We will not work out the full distribution, but only the first moment
$\expct[\rho]{\tilde{v}}^x = \expct[\rho]{v}^x -\efst{v}^x$.
Since $\rho$ is a distribution only over $x$, this requires some
explanation.
In accordance with the preceding discussion,
the expectation with respect to $\rho(x)$ of a function $g$ of both $x$ and $v$
should be read as
 \begin{equation}
   \label{eq:x-v-expectations}
 \expct[\rho]{g(x,v)} = \expct[\rho\ltimes\rho_{\text{fst}}]{P^T g},  
 \end{equation}
while for $g$ a function on $T_xQ$, $\expct[\rho]{g}^x$ is the corresponding
distribution conditioned on position being $x$.
The $P^T$ operator in (\ref{eq:x-v-expectations}) means
that the velocity has had a chance to equilibrate with the ``environment''
described by the position distribution $\rho$.
A little reflection should convince one that the current density is
related to the instantaneous velocity expectation via
\begin{equation}
\label{eq:J-v}
  J[\rho] = \expct[\rho]{v}\, \rho.
\end{equation}
(For one-stage contraction; two stages is a bit more complicated.)
The only real question is whether the framework of Section \ref{sec:contraction}
is able to deliver this conclusion. In order to calculate the first moment
$\expct[\rho]{\tilde{v}}^x$ of the conditional mean instantaneous velocity,
it will be enough to compute
\begin{equation}
  \label{eq:polarization-test}
\expct[\rho\ltimes\rho_{\text{fst}}]{ P^T \tilde{v}^i f },
\end{equation}
for an arbitrary test function $f(x)$ of position only.
Now, recall (\ref{eq:Duhamel-Dyson})
\begin{equation}
  \nonumber
P^T \simeq \Pfst^T + \int_0^T \Pfst^{T-t} (v\cdot\partial_x) \Pfst^{t} \, dt.
\end{equation}
$\Pfst$ is oblivious to $f$, so the first term gives
a contribution $\int \rho(x) f(x)\efst{\tilde{v}^i}^x \, d\Omega = 0$
to (\ref{eq:polarization-test}).
The next term is more interesting. Since $T$ is infinitely
large compared to the fast relaxation time, we obtain
(with an integration-by-parts in the final step)
\begin{align}
  \label{eq:polarization-test-calc}
  \expct[\rho\ltimes\rho_{\text{fst}}]{ P^T \tilde{v}^i f }
& = 
  \expct[\rho]{
  \int_0^\infty dt \efst{ v^j \partial_j[ \tilde{v}^i(t) f] }^x
      }
  \nonumber
  \\
   &= 
     \int \rho \left\{ (\con{u}^i - \efst{v}^i) f
     - \con{D}^{ij} \frac{\partial f}{\partial x^j} \right\}
     \, d\Omega
     \nonumber \\
   &= 
     \int f \left\{ (\con{u}^i - \efst{v}^i) 
     - \frac{1}{\rho}\DIV (\evec{i}\cdot\con{D}\rho) \right\} \, \rho d\Omega.
     \nonumber
\end{align}
Since $f$ is an arbitrary test function,
we conclude that
\begin{equation}
\expct[\rho]{v^i}^\cdot =
\con{u}^i - \frac{1}{\rho}\DIV (\evec{i}\cdot\con{D}\rho).
\end{equation}
Comparing this to Eq. (\ref{eq:current-density}) for the probability current
density, one see that (\ref{eq:J-v}) is indeed verified.

All this prompts one to ask which kind of measurement would be expected to
yield the drift velocity $\con{u}$, which the instantaneous velocity $\expct[\rho]{v}$.
The calculations used here to find them both involved evolving the system for a time
$T \gg \tau$, however, that evolution served very different purposes in the two cases.
Suppose the system is started (or found) at $x$. Then, $\con{u}$ is the expected
average velocity of the system over the subsequent time $T \gg \tau$.
If now, the system has a probability density $\rho$ (perhaps we have calculated this
for a time $T' \gg T$ via the Fokker-Planck equation from an initial point mass at $x$),
then our expectation of the {\it instantaneous} velocity of the system, regardless of
its actual position, is $\expct[\rho]{v}$. ``Instantaneous'' here means that the measurement
is done over a time small (or at worst, comparable to) $\tau$.

\section{Conclusion}
\label{sec:conclusion}

By means of a ``Heisenberg picture'' approach,
elimination of fast variables was carried out in general for {velocity models},
resulting in a general expression for the infinitesimal generator $\con{\Gen}$,
equivalently, the drift-diffusion field $(\con{u},\con{D})$, of the contracted
drift-diffusion process on position space in terms of time correlation functions in
fast-equilibrium.
Compared to Chapman-Enskog type expansions, which work in the ``Schr\"odinger picture'',
the method seems to be simpler and more automatically general.
Information about the distribution of fast variables, which comes naturally with
Chapman-Enskog expansions, is also available in the current approach, if desired
(Sec. \ref{sec:instantaneous}).
Examination of the subclass of quasilinear models showed that drift, other than
that already present in fast-equilbrium, traces to derivatives of either
velocity decay ($\Lambda$) or the dependence of velocity on fast variables ($\sigma$).
The first of these is exemplified by the humble brownian particle in a viscosity
gradient, and the second by the active particle in a fuel concentration gradient.

\begin{acknowledgments}
I thank Prof. Vincent H. Crespi for critical reading and suggestions.
This work was funded by the Penn State MRSEC, Center for Nanoscale Science, under 
award National Science Foundation DMR-1420620.
\end{acknowledgments}
%

\end{document}